# Tunable Collective Excitations in Epitaxial Perovskite Nickelates

Mengxia Sun,[1,#] Xu He,[2,#] Mingyao Chen,[1] Chi Sin Tang,[1,3,*] Xiongfang Liu,[1] Liang Dai,[1] Jishan Liu,[4] Zhigang Zeng,[1] Shuo Sun,[1] Mark B.H. Breese,[3,5] Chuanbing Cai,[1] Le Wang,[6,*] Yingge Du,[6] Andrew T. S. Wee,[5,7] Xinmao Yin[1,*]

[1]Shanghai Key Laboratory of High Temperature Superconductors, Institute for Quantum Science and Technology, Department of Physics, Shanghai University, Shanghai 200444, China

[2]Theoretical Materials Physics, Q-MAT, CESAM, Université de Liège, B-4000 Liège, Belgium

[3]Singapore Synchrotron Light Source, National University of Singapore, Singapore 117603, Singapore

[4]State Key Laboratory of Functional Materials for Informatics, Shanghai Institute of Microsystem and Information Technology, Chinese Academy of Sciences, Shanghai 200050, China

[5]Department of Physics, Faculty of Science, National University of Singapore, Singapore 117542, Singapore

[6]Physical and Computational Sciences Directorate, Pacific Northwest National Laboratory, Richland, WA 99354, USA

[7]Centre for Advanced 2D Materials and Graphene Research, National University of Singapore, Singapore 117546, Singapore

*Corresponding author. Email: slscst@nus.edu.sg (C.S.T.); le.wang@pnnl.gov (L.W.); yinxinmao@shu.edu.cn (X.Y.)
# These authors contributed equally to this work.

**Abstract**

The formation of plasmons through the collective excitation of charge density has generated intense discussions, offering insights to fundamental sciences and potential applications. While the underlying physical principles have been well-established, the effects of many-body interactions and orbital hybridization on plasmonic dynamics remain understudied. In this work, we present the observation of conventional metallic and correlated plasmons in epitaxial $La_{1-x}Sr_xNiO_3$ (LSNO) films with varying Sr doping concentrations (x = 0, 0.125, 0.25), unveiling their intriguing evolution. Unlike samples at other doping concentrations, the x = 0.125 intermediate doping sample does not exhibit the correlated plasmons despite showing high optical conductivity. Through a comprehensive experimental investigation using spectroscopic ellipsometry and X-ray absorption spectroscopy, the O2$p$-Ni3$d$ orbital hybridization for LSNO with a doping concentration of $x$ = 0.125 is found to be significantly enhanced, alongside a considerable weakening of its effective correlation $U^*$. These factors account for the absence of correlated plasmons and the high optical conductivity observed in LSNO (0.125). Our results underscore the profound impact of orbital hybridization on the electronic structure and the formation of plasmon in strongly-correlated systems. This in turn suggest that LSNO could serve as a promising alternative material in optoelectronic devices.

# 1. Introduction

The complex interactions between the lattice, charge, spin, and orbital degrees of freedom regulates the electronic and magnetic structures of strongly-correlated systems, resulting in intriguing physical phenomena including superconductivity, giant magnetoresistance effect, and metal-insulator transition.[1-4] The impact of many-body interactions and orbital hybridizations on the electronic properties remains a highly sought-after research domain in perovskite structure transition metal oxides.[5-7] Meanwhile, collective charge dynamics in the form of plasmon excitations are intimately linked to the electronic characteristic of many materials.[8-10]

While conventional metallic plasmons have been observed in multiple strongly-correlated transition metal oxides,[11-13] note that theoretical studies have revealed the interaction of both short- and long-range correlations in Mott insulating systems can produce unconventional plasmons known as 'correlated plasmons' which possess multiple ordered photon energies.[14] These correlated plasmons have been experimentally observed in niobates, cuprates, and topological materials such as $Bi_2Se_3$, where charge-spin coupling plays an important role in the formation of correlated plasmons.[8, 10, 15] Meanwhile, orbital hybridization between the metallic $3d$ and oxygen $2p$ orbitals have been pivotal in accounting for multiple quantum phenomena in strongly-correlated systems,[16-18] their role in mediating the formation of plasmons remains largely an unchartered research frontier that requires further investigation. Moreover, as correlated plasmons are intimately related to long-range charge correlations,[10, 14] the detection of this unconventional plasmons holds the potential to reveal valuable insights to the underlying mechanisms that result in their formation.

Rare earth nickelates have a typical perovskite structure, $ABO_3$ (A and B are cations of different volumes), where A-site is a rare earth atom surrounded by eight $NiO_6$ octahedrons and have garnered significant attention due to their unique electronic structures and rich phase diagram.[19-21] Among them, $LaNiO_3$ stands out with its metallic behavior and bifunctional catalytic activity.[21-24] Through the process of Sr-doping and other treatment techniques, the observation of charge density waves and the emergence of superconductivity have raised further interest amongst the family of nickelates.[25-28] Meanwhile, $LaNiO_3$ is a typical correlated metal located near the metal-insulator transition boundary. By substituting the A-site with other rare earth ions,[29] or by confinement,[30] it is possible to induce a transition to a Mott insulator. Scientific investigation into the formation and regulation of unconventional plasmons is essential for gaining a deeper understanding of electronic correlations. Here, we present our observations and insights into the optical conductivity and unique paired energy correlated plasmons in $La_{1-x}Sr_xNiO_3$ (LSNO) systems with varying doping concentration ($x = 0$ (LNO); x = 0.125 (LSNO (0.125)); x = 0.25 (LSNO (0.25)). We found that conventional metallic plasmons and correlated plasmons coexist in LNO, but the correlated plasmons disappear in LSNO (0.125). At higher Sr doping concentration in LSNO (0.25), the correlated plasmons re-emerge. Additionally, LSNO (0.125) exhibits significantly higher optical conductivity compared to LNO and LSNO (0.25). Besides the hole doping effect, the increasing of the Sr concentration can also affect the inter-site hybridization and the effective charge correlation. A series of spectroscopic ellipsometry (SE) and X-ray absorption spectroscopy (XAS) measurements substantiated demonstrate that the transitions between different plasmonic states and the anomalous optical conductivity can be attributed to variations in the O2$p$-Ni3$d$ orbital hybridization and effective interaction $U^*$ (comprising the long-range Coulomb variable interactive components). Analysis of the effective interaction, $U^*$, suggests that the orbital hybridization effects may indirectly influence the long-range charge correlations in the LSNO system. Consequently, orbital hybridizations impact material plasmons excitation, regulate long-range electron correlation, and enhance optical conductivity fluctuation. These findings have promising implications for plasmonic-based device engineering research and our understanding of electronic correlations in strongly correlated systems.

## 2. Results and Discussion

### 2.1 Optical conductivity of LSNO epitaxial thin films

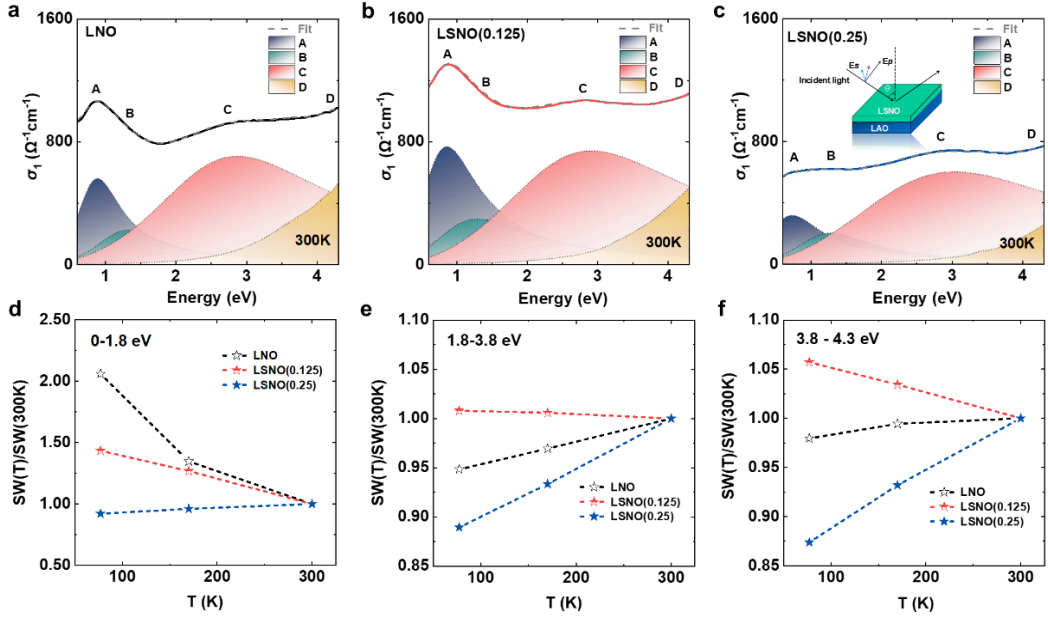

**Figure 1.** (a) Real component of the optical conductivity, $\sigma_1$, of LNO, (b) LSNO (0.125), (c) LSNO (0.25) films at room temperature. Experimental spectra fitted by Drude-Lorentz peaks (shown as light-colored peak) and visual guides to identify positions of peaks A, B, C, and D in $\sigma_1$ spectra and the gray dotted line is the result of the fit (Inset: Schematic illustration of spectroscopic ellipsometry experiment). Spectral weight (defined as SW (T) / SW (300K)) of LNO, LSNO (0.125), and LSNO (0.25) in different spectral regions: (d) SW1 (0–1.8 eV), (e) SW2 (1.8–3.8 eV), (f) SW3 (3.8-4.3 eV). T is temperature (Kelvin). The magnitude of $\sigma_1$ at photon energy range below ~0.6 eV is estimated using a linear interpolation based on the Drude model. Note that while the approximation may deviate from the actual quantity, the estimation remains accurate to describe the electronic characteristics at low energy levels.

20 unit cells (u.c.) LSNO thin films with different Sr doping levels ($x$ = 0, 0.125, 0.25) were grown on (001)-oriented LaAlO$_3$ (LAO) single crystal using oxygen plasma–assisted molecular beam epitaxy (OPA-MBE).[23] As shown in Figure S1, the diffraction peaks of the LSNO films are all located directly below the peaks of the substrate LAO in the reciprocal space mapping (RSM) diagrams without observing other diffraction peaks. This suggests that the LSNO films are of high crystalline quality, epitaxially grown under complete strain and devoid of any secondary phases, making them well-suited for subsequent spectroscopy investigations. Moreover, in-plane transport measurements confirm that all three samples demonstrate metallic behavior throughout the entire temperature range between 2K and 300 K (Figure S2), with the electrical resistivity increasing with increasing doping concentration, *x*. Hence, as temperature decreases, the LSNO samples do not undergo metal-insulator transition (MIT), which is unlike other nickelates.[31, 32]

In order to analyze the changes of optical structures with LSNO system, optical conductivity, $\sigma_1$, (Figure 1a-c), was determined from the spectroscopic ellipsometry (SE) measurements. To accurately analyze the optical features in each spectrum, the optical analysis software of public domain REFFIT has been utilized where the Drude-Lorentz model is employed to account for both the delocalized and localized charges in the respective systems.[33] The LSNO films at different doping concentrations exhibited generally similar optical characteristics, featuring distinct optical features labeled as A, B, C, and D (Figure. 1a-c). By analyzing the spectral properties with those reported in previous studies[34, 35] and our calculations (See supplementary materials), these optical features can be attributed to various inter-band transitions. The feature A of $\sigma_1$ spectra located at ~0.89 eV in the LNO film corresponds to the transition from Ni($t_{2g}$) orbital to the Fermi surface while feature B (~1.32 eV) corresponds to transitions from Ni($t_{2g}$) to the unoccupied Ni($e_g$) orbital. The broad feature denoted C (~2.87 eV) is associated with transitions between the O2$p$ and Ni($e_g$) orbitals. Meanwhile feature D, with its centre located beyond the instrument measurement range, may be ascribed to the transition from the O2$p$ orbital to antibonding Ni($e_g^A$) orbital or the bonding Ni($e_g^B$) orbital to Ni($e_g$) orbital. The exact location of this optical feature is beyond our current measurements' capabilities. Similar analysis has also been performed for the other samples and the electronic band transition diagram is obtained (Figure S4). These findings shed that Ni($e_g$) orbitals is away from the Fermi surface in response to increasing Sr doping, contributing to a deeper understanding of the optical properties of the materials under investigation.

In the photon energy range of our measurements (Figure 1b), the intensity of the $\sigma_1$ spectra of LSNO (0.125) is distinctly stronger compared to other samples. To comprehensively assess the electronic behavior within the system, we estimated the magnitude of $\sigma_1$ for LSNO at 0 eV by employing linear interpolation based on the transport results depicted in Figure S1.[36] This is approximated using the Drude model, while this approximation may deviate slightly from the actual values, it still in good agreement with the deduced electronic characteristics at low photon energy.[36, 37] A partial spectral weight (SW) integral of an energy region can be defined as $SW = \int_{\omega_1}^{\omega_2} \sigma_1(\omega)d\omega$. According to the previous analysis of electronic structure, $\sigma_1$ is divided into three regions: below 1.8 eV (low-energy region, SW1 for spectral weights), 1.8 eV-3.8 eV (middle-energy region, SW2 for spectral weights), and 3.8 -4.3 eV (high-energy region, SW3 for spectral weights). In low energy region (Figure 1d), the SW1 of LNO is the strongest because of its abundance of free charges. Notably, in middle-energy and high-energy regions (Figures 1e –f) this observation aligns with the findings in Figure 1b, where the SW appears increased in LSNO (0.125). Moreover, the SW of LNO and LSNO (0.125) exhibit similar temperature-dependent trends in the low energy region while LNO and LSNO (0.25) have the same temperature-dependent response in the high energy region. It's worth noting that the SW of LSNO (0.125) and LSNO (0.25) exhibit opposite temperature-dependent trends in all energy region, which imply the different electron correlation characteristics. And the reasons will be analyzed later. Moreover, as demonstrated in Figure S5, the feature peak positions of optical conductivity remain generally unchanged with temperature. This implies that, unlike other rare earth nickelates,[38, 39] the correlation strength in the LSNO system is relatively insensitive to changes in temperature.

**2.2 Conventional and Correlated Plasmons of LSNO epitaxial films**

An analysis of the dielectric function, $\varepsilon(\omega) = \varepsilon_1(\omega) + i\varepsilon_2(\omega)$ (where $\omega$ is the photon angular frequency) is performed for each sample, along with their corresponding loss function (LF, where $LF = Im(\frac{-1}{\varepsilon(\omega)}) = \frac{\varepsilon_2(\omega)}{\varepsilon_1^2(\omega)+\varepsilon_2^2(\omega)}$) as displayed in Figure 2. This set of analysis is essential to provide additional insights into the plasmonic behaviors of LSNO epitaxial films and how they evolve with varying doping concentration.

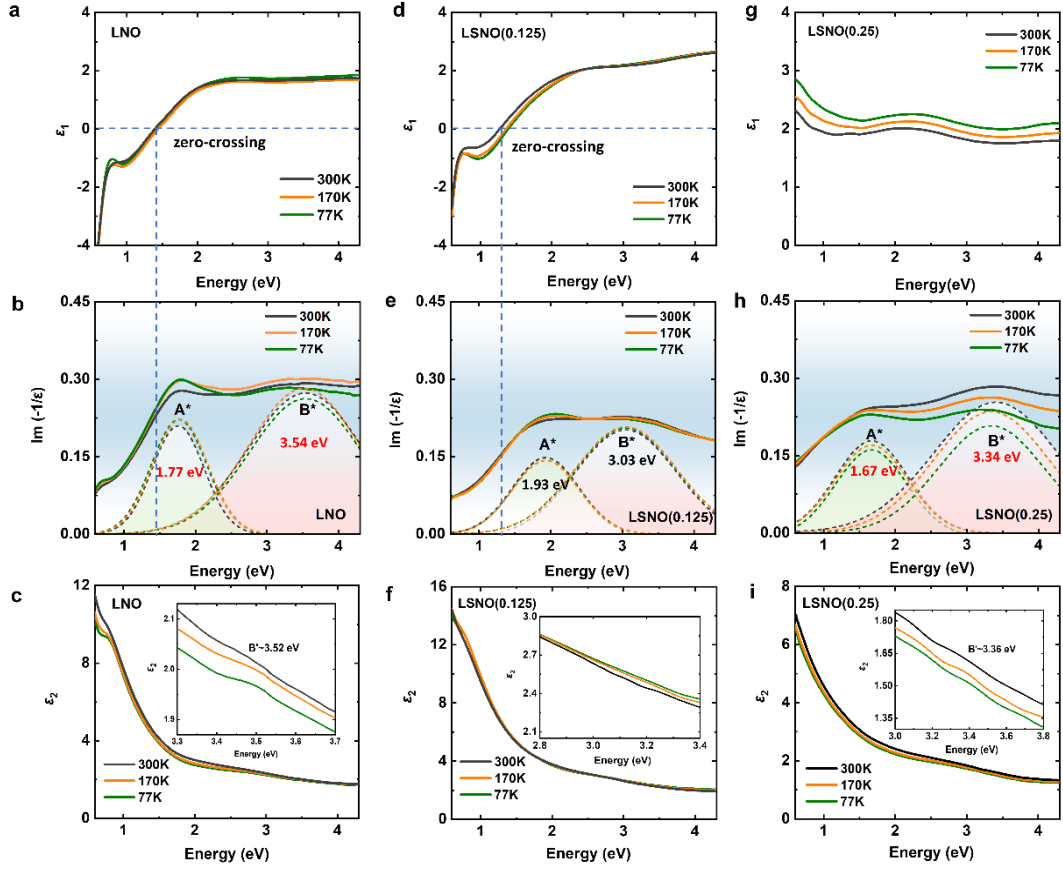

**Figure 2**. (a, d, g) Real ($\varepsilon_1$) components of LNO films (a) and LSNO (0.125) films (d) and LSNO (0.25) films (g) at different temperature. (b, e, h) Loss-function spectra of LNO films (b) and LSNO (0.125) films (e) and LSNO (0.25) films (h) at 300K, 170K, 77K. The peaks in dashed lines are Voigt profile fitted peaks. (c, f, i) Imaginary ($\varepsilon_2$) components of (c) LNO films (f) and LSNO films (i) and LSNO (0.25) films at different temperature. Light blue dashed lines serve as visual guides in identifying positions of $\varepsilon_1$ zero-crossing and the silver dotted line is the result of the fit at 300K.

Two distinct peaks labelled A* and B* have been observed in the LF spectra of the respective samples. These peaks are absent from their corresponding $\sigma_1$ and $\varepsilon_1$ spectra. As shown in Figure 2a and 2b, peak A* (~1.77 eV) in the LF spectra of LNO is approximately in similar energy position with the zero-crossing of the corresponding $\varepsilon_1$ spectrum at ~1.40 eV and move indistinctively with temperature. This alignment is a signature of conventional metallic plasmons, which are present in the metallic state of LNO.[10, 12, 40] In the case of LSNO (0.125), the conventional metallic plasmons remain with peak A* (~1.93 eV) in the LF spectrum coinciding with the corresponding $\varepsilon_1$ zero-crossing at ~1.28 eV. The disparity in the positions of the $\varepsilon_1$ zero-crossing and the corresponding LF peaks can be attributed to the onset of free-electron scattering within the system.[8, 10] For the LSNO (0.25) sample, the $\varepsilon_1$ spectrum no longer exhibits any zero-crossing and corresponding with the reduction of metallic properties. This shows that with the increase in the concentration of Sr doping, there is a corresponding decrease in the free carrier concentration in the LSNO system, which is consistent with the transport results.

In addition to peak A*, peak B* is also present in the LF spectra of the respective LSNO samples at a higher photon energy (Figure 2b, 2e and 2h). Particularly for LNO and LSNO (0.25), there is a two-fold energy relationship between peaks A* and B*(LNO: peak A*~1.77 eV, peak B*~3.54 eV; LSNO (0.25): peak A*~1.67eV, peak B*~3.34 eV). However, such a two-fold energy relation between peaks A*(~1.93 eV) and B*(~3.03 eV) is absent from LSNO (0.125). Combining the optical features observed in the $\sigma_1$ spectra with peaks A* and B* in LF spectra, it becomes evident that these peaks do not originate from inter-band transitions, as seen in their $\sigma_1$ spectra. This suggests that, apart from the conventional metallic plasmon behavior represented by peak A* of LNO and LSNO (0.125) films, there may be other plasmonic phenomena at play in LSNO system with the two-fold relationship between features A* and B* in LF spectra. Notably, similar experimental observations concerning correlated plasmons in Mott-like insulating oxides have been reported.[8, 10] These findings provide additional impetus for further investigation to the plasmonic behavior exhibited in LSNO samples.

As shown in Figure 2b and 2c, an important observation can be noticed in the $\varepsilon_2$ and LF spectra of the LSNO films. For LNO, in addition to the distinct Drude peak in the low energy region, characteristic peak B'(~3.52eV) is also visible in the $\varepsilon_2$ spectra. Particularly for peak B', it is located at almost the same energy position with the feature B*(~3.54 eV) of the LF spectrum. Previous studies have shown that the existence of both peaks in $\varepsilon_2$ and LF spectra suggest a strong coupling between the optical and plasmonic excitations of LNO and is indicative of correlated plasmons.[8, 41] This implies that peak B* is the correlated plasmons excitation peak in LNO. Similarly, in LSNO (0.25), peak B*(~3.34 eV) and peak B'(~3.36 eV) were observed in the same position in $\varepsilon_2$ spectrum and LF spectrum, suggesting the presence of correlated plasmons as well.

The observation of correlated plasmon has direct relevance in accounting for the presence of correlated plasmons. To elucidate the two-fold relationship present in the LF spectra for both LSNO (0.25) and LNO, theoretical studies have suggested that the dispersion relation sufficiently changes and at the strong interaction strength,[42] a pair of plasmonic peaks may arise in the LF spectrum, with one peak at energy $U^*$ and another at quantization energy $\sim U^*/2$, where $U^*$ denotes the effective interaction.[14] The peak at $U^*$ stems from collective excitations in the Mott correlated bands, while the peak at $U^*/2$ can be attributed to the excitation between the quasi-particle band and the Hubbard bands.[14] This may indicate peak A* is correlated plasmons excitations in both LSNO (0.25) and LNO. Specifically, peak B* at ~3.34 eV in the LF spectra of LSNO (0.25) (Figure 2h) is equivalent to $U^*$ while Peak A*($U^*/2$) located half the energy of peak B* at ~1.67eV is the half-energy mode of the correlated plasmon associated with peak B*. This two-fold relationship between features A*(~1.77eV) and B*(~3.54 eV) is also present in LNO. This observation may suggest the coexistence of both conventional metallic and correlated plasmons within the LNO system and peaks A* are both conventional metallic plasmon and correlated plasmons excitation peaks, which coincide or are indistinguishable.

The 2-fold energy relationship first proposed in Ref. 14 is with dynamical-mean-field theory (DMFT) which take into consideration the charge correlation effect in a simplified half-filled single-orbital extended Hubbard model. The DMFT calculation for LNO $e_g$ states show very similar three bands featuring the correlated metal: a quasi-particle band at the Fermi energy and the two Hubbard bands below and above the Fermi energy, respectively with a moderate Hubbard energy, and can only become insulating with a very large Hubbard energy.[43] The similar results can be found in XAS experiments in thin films.[44]

Moreover, it is noteworthy that the presence of the correlated plasmons persists across the entire temperature range we measured in both LNO and LSNO (0.25), as evidenced by the $\varepsilon_2$ spectra and presence of the 2-fold relationship between A* and B* in the LF spectra for both systems, as shown in Figures 2(b) and (h). It is essential to highlight that the correlated plasmons exhibits significant variations with decreasing temperature, resulting in a notable reduction in plasmon peak intensity (Figs. 2(b) and (h)). In contrast, the excitation of conventional metallic plasmons (e.g., A * of LSNO

(0.125)) remains largely unaffected with decreasing temperature. This disparity can be attributed to the unique behavior of correlated plasmons, which do not follow the conventional pattern associated with the zero-crossing of $\varepsilon_1$, as seen in the traditional metallic systems (absent in the case of correlated plasmons).[8, 10] This phenomenon is particularly pronounced in LSNO (0.25), where the correlated plasmons originate from the oscillatory dynamics of localized correlated electrons rather than the free charges in conventional metallic systems. The contrast between the $\varepsilon_1$ spectra of LSNO (0.125) and LSNO (0.25) serves as a clear illustration of this distinction. The presence of correlated plasmons in LSNO adds to the complexity and richness of the electronic behavior in this intriguing material system.

Furthermore, through the comparison of $\varepsilon_1$ spectrum and LF spectrum and the absence of correlated plasmons in LSNO (0.125) can be attributed to the effective interaction, $U^*$, having been weakened below a critical threshold energy known as $U_c^*$. Below this threshold, it is no longer favourable for the formation of correlated plasmons because electronic correlation become weaker.[14] Consequently, the broad plasmon peak B* undergoes a redshift to ~3.03 eV emerges observed in medium-high energy region in LSNO (0.125), but note that this is not correlated plasmons without coupling between the optical and plasmonic excitations. The unique plasmon behavior in LSNO systems provides valuable insights into electronic correlations behaviors, which illustrate the reduction of the collective oscillations of the associated electrons and thus the correlated plasmons ultimately vanish in LSNO (0.125).

**2.3 X-ray Absorption Spectra of LSNO epitaxial films**

To gain further insights into how the electronic correlations and structure of the LSNO system evolve with Sr doping, detailed X-ray absorption spectroscopic (XAS) measurements are conducted on the respective samples. Figure 3a display the Ni $L_3$-edges of the respective LSNO samples at grazing incidence (GI). Besides the prominent La $M_4$-edge peak (~853.4eV) belonging to the La components in both the LSNO film and LAO substrate, the Ni $L_3$-edge feature appears in the form of feature $a$, along with a shoulder denoted as $a'$. These characteristics provide valuable information about the local Ni environment.[45-49]

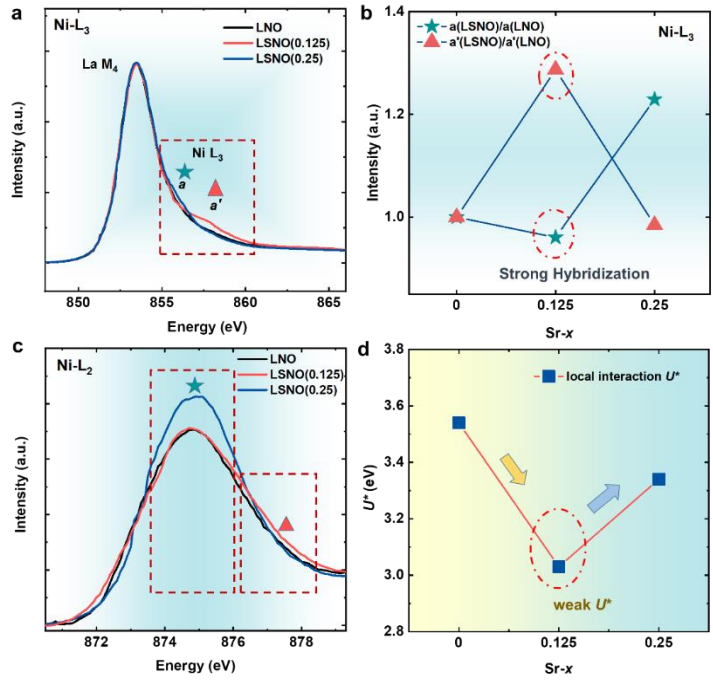

Figure 3. (a) Grazing-incidence (GI) XAS at the Ni $L_3$ edge for LNO films (black), LSNO (0.125 films (red) and LSNO (0.25) films (blue) at room temperature. (b) Voigt profile fitting conducted results for the intensity of respective Ni $L_3$ edge features a and a'. Showing the relative intensity of features peak $a$ ($\frac{a(LSNO)}{a(LNO)}$) and peak $a'$ ($\frac{a'(LSNO)}{a'(LNO)}$). To prevent interference from the La $M_4$ edge feature, a consistent half-peak width and the peak position for the La $M_4$ edge is kept in place during the fitting process. (c) Normalized Ni $L_2$ edge for LNO films (black), LSNO (0.125) films (red) and LSNO (0.25) films (blue) at room temperature. (d) Effective interaction $U^*$ of LSNO (x = 0, 0.125, 0.25) films with varying doping concentration. It is extracted from position of peak B* of LF spectrum in Figure 2. Red dotted circles serve as visual guides in identifying extreme value position of hybridization and effective interactions $U^*$ with different Doping concentration in LSNO.

To examine the evolution of the intensity and position of features $a$ and $a'$ between LNO, LSNO (0.125) and LSNO (0.25), Voigt profile fitting is performed for the respective Ni $L_3$-edge features (see supplementary materials Figure S7) with the results displayed in Figure 3b. Across all three samples, the peak positions remain consistent. However, for LSNO (0.125), feature $a$ has the lowest intensity while the intensity of feature $a'$ maximises. To gain a deeper understanding of the orbital behavior, charge transfer multiplet calculations for $Ni^{3+}$ in the D4$h$-symmetry were further conducted using CTM4XAS (Figure S8).[50] The outcomes of these calculations indicate that as the splitting parameter 10Dq increases, peak $a'$ becomes more pronounced while peak $a$ weakens. Previous theoretical calculations on LNO with a mixture of configurations $3d^7$, $3d^8\underline{L}$, $3d^9\underline{L}^2$ (where $\underline{L}$ denotes a ligand hole) in the ground state suggest that the variation in intensity of features $a$ and $a'$ is corresponds to changes in the Ni-O hybridization strength instead of changes to the Ni oxidation states.[45, 48, 49] Specifically, peak $a$ arises due to the $2p^63d^7$—$2p^53d^8$ transition, while peak $a'$ comes from the $2p^63d^8\underline{L}$—$2p^53d^9\underline{L}$ transition.[50] Meanwhile, previous theoretical and experimental studies have indicated that a higher intensity of feature $a'$ than $a$ indicates a stronger O2$p$-Ni3$d$ hybridization strength. This suggests a direct correlation with an increased contribution of the $3d^8\underline{L}$ orbitals.[46, 49] Hence, it becomes evident that the intensity of the feature $a'$ is closely tied to the strength of the O2$p$-Ni3$d$ hybridization Where stronger hybridization leads to an increased intensity of feature $a'$. As shown in Figure 3b, the intensity of feature $a'$ which maximizes for LSNO (0.125) indicates that the O2$p$-Ni3$d$ hybridization is strongest for the LSNO sample at this doping concentration, it is weakest for LSNO (0.25).

The Ni $L_2$ edge, stemming from transitions from Ni $2p_{1/2}$ to Ni $3d$, is unaffected by the La $M_4$ edge. Hence, it serves as an important X-ray absorption feature for analyzing the strength of Ni-O orbital hybridization.[47, 51, 52] As depicted in Figure 3(c), both the primary and the shoulder peak features are evident in the individually normalized Ni $L_2$ edge spectra of the LSNO system, with the peak positions exhibiting stability. Notably, the LSNO (0.125) sample exhibits the weakest primary peak (~875.1 eV) and the strongest shoulder peak (~877.2 eV), while LSNO (0.25) displays the strongest primary peak and the weakest shoulder feature. This trend is identical with the behavior of the Ni $L_3$ edge feature. Therefore, the consistency between the Ni $L_{2,3}$ edge analyses reinforces our conclusion that Ni-O orbital hybridization is strongest in LSNO (0.125) and weakest in LSNO (0.25).

Furthermore, by comparing the effective interaction, $U^*$ (Figure 3c), it can be concluded that $U^*$ decreases with increasing doping concentration between LNO and LSNO (0.125). But as doping concentration continues to increase from LSNO (0.125) to LSNO (0.25), $U^*$ rises again. Therefore, $U^*$ is minimized for LSNO (0.125). The O2$p$-Ni3$d$ orbital hybridization strength shows the opposite trend where it is maximized for LSNO (0.125) but is weakest for LSNO (0.25). Collectively, these trends suggest that orbital hybridization may indirectly influence the effective interaction strength, $U^*$. This, in turn, has the potential to modulate long-range interactions within the system, as $U^*$ is a composite parameter that encompasses both long-range $V(q)$.

## 2.4 DFT Calculates the Electronic Structure of LSNO Epitaxial Films

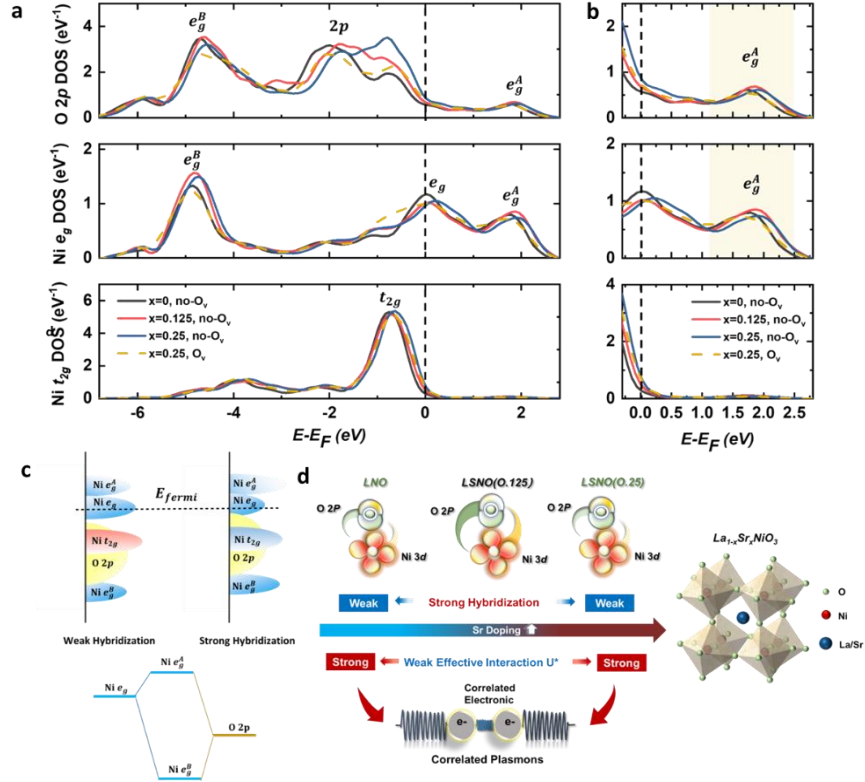

**Figure 4.** (a) Density of states (DOS) of the respective constituent (Ni $e_g$, Ni $t_{2g}$ and O $2p$) orbitals for different doping concentration of x = 0, 0.125, 0.25 in LSNO without oxygen vacancy (no-$O_v$). For x = 0.25, the presence of an oxygen vacancy ($O_v$) case is considered, as depicted by the dashed line. (b) Enlarged view near the Fermi surface in Figure 4a. (c) Schematic of the electronic structure in LSNO. Black dashed lines serve as visual guides in identifying positions of Fermi surface. The $t_{2g}$ bands can be further decomposed into bonding and anti-bonding states but we do not do that for simplicity, as they are fully occupied. (d) Illustration of the LSNO crystal structure and the underlying mechanism governing correlated plasmons. The co-regulation of plasmon appearance and vanished is governed by the concerted influence of Ni-O hybridization and the effective interaction parameter $U^*$.

First principles calculations are carried out based on density functional theory (DFT) method without the Hubbard U on LSNO. Figure 4a displays the density of states (DOS) projected in the O 2$p$, Ni $e_g$ and Ni $t_{2g}$ orbitals for the LSNO structures with Sr-doping of $x = 0$, $x = 0.125$, $x = 0.25$, respectively. Within the energy range from -5 to 2 eV, distinct peaks emerge in both the O 2$p$ and Ni $e_g$ states. These peaks can be attributed to the formation of bonding ($e_g^B$) and antibonding ($e_g^A$) states resulting from the hybridization of these two orbitals. Notably, there is another Ni $e_g$ peak positioned at the Fermi level, however, it does not show substantial hybridization with O 2$p$ orbitals, signifying its non-hybridized nature. The non-hybridized O 2$p$ states can be found around -2 eV. The $t_{2g}$ center is about -1 eV below the Fermi energy.

Furthermore, the DOS are observed to shift towards a higher energy region with respect to the Fermi level with increasing Sr doping concentration. This shift can be attributed to the decrease in occupied states due to an increasing hole doping. As a hole-doped system, further attention will be placed on the behavior of the hole carriers above the Fermi level.[53, 54] As shown in Figure 4b, the most notable change observed with varying doping concentration ($x$) is the maximization of the DOS for $e_g^A$ states of both Ni and O for LSNO (0.125). This is followed by a subsequent decrease with further hole doping. Note also that the strength of feature $e_g^A$ in LSNO (0.25) is lower than that in LNO, indicating a reversed trend when compared to Ni $e_g$ states. Moreover, according to the DFT calculations, the formation energy of oxygen vacancies, $O_V$, decreases with increasing Sr doping concentration, making it easier for these $O_V$ to form. Therefore, for the case of LSNO (0.25), oxygen vacancies are considered to have a greater influence and this leads to the significant reduction in $e_g^A$ after the introduction after its introduction.

According to the experimental results, a schematic could be plotted to visualize the electronic structure of LSNO as shown in Figure 4c. From left to center panel, the hybridization between the O and the Ni states are enhanced, and the more states goes from the non-hybridized to the hybridized, which is accompanied by the charge transfer from O 2$p$ to the Ni-O bonding state. In the language of the Ni electronic configuration, it goes from Ni 3$d^7$ to 3$d^8\underline{L}$ or even 3$d^9\underline{L}^2$, where $\underline{L}$ is the ligand hole.

The presence of correlated plasmons is particularly pronounce in LSNO (0.25) which registers the weakest Ni-O hybridization. This is followed by LNO. Conversely, the correlated plasmons vanish in LSNO (0.125) where Ni-O hybridization is the strongest. Meanwhile, the presence of conventional metallic plasmons follows an opposite trend where it vanishes in LSNO (0.25) where orbital hybridization is at its weakest. As discussed earlier, occupation, orbital hybridization and effective interactions, $U^*$, have strong influences on each other. Hence, we propose that the maximized Ni$3d$–O$2p$ orbital hybridization, coupled with the diminishing $U^*$ in LSNO (0.125) play a significant role in influencing correlated charge excitation (Figure 4d). Meanwhile, as shown in Figure 1d-1f, we observed the variations in spectral weights with doping, the SW of LSNO (0.125) and LSNO (0.25) exhibit contrasting trends with temperature. However, the spectral weights of LNO shows the same trend as LSNO (0.25) at SW2 and SW3 region, but the same trend as that of LSNO (0.125) in SW1 region. Since the spectral weights are conserved, this implies that correlated plasmons transfer low-energy spectral weights towards the high-energy region, whereas metal plasmons induce a transfer of high-energy spectral weights towards the low-energy region. This trend suggests the augmentation in optical conductivity in LSNO (0.125), specifically in the SW2 and SW3 regions.

An interesting observation of correlated plasmons have been made in LSNO films, a strongly-correlated Mott insulating system. A deeper understanding of plasma properties holds paramount importance in their applications in photovoltaic technology and plasma solar cells. Notably, correlated plasmons observed in strongly-correlated systems such as LSNO are fundamentally different from the conventional ones found in metallic systems such as gold,[55, 56] graphene,[57] and VO$_2$.[58, 59] In which case they manifest themselves in multiple plasmon frequencies within the visible–ultraviolet range. The multifrequency characteristic of LSNO enhances optical modulation flexibility, adapting to diverse wavelength ranges. Aligned with correlated plasmons observed in copper oxide-based high-temperature superconductors,[10, 41] LNO exhibits a unique feature: simultaneous excitation of conventional and correlated plasmons in same position of LF spectra—an unprecedented phenomenon. This supports the efficient coupling of correlated plasmons with free-space photons, enhancing metallic plasmon excitation. As a perovskite oxide, LNO shows promise as an oxygen electrocatalyst for renewable energy storage and as bottom electrodes. Our research emphasizes the significant role of Ni-O orbital hybridization in shaping correlated plasmon formation, dissipation, and increased optical conductivity in LSNO samples. This distinct attribute positions LSNO as a compelling alternative in optoelectronic devices.

## 3. Conclusion

In summary, the evolution of conventional and correlated plasmons and the optical conductivity behavior in the LSNO spectra at different doping concentrations have been discussed and analyzed based on the experimental results elucidated from SE and XAS measurements. Specifically, we have observed the presence of correlated plasmons in both LNO and highly-doped LSNO (0.25), whereas the conventional plasmons, accompanied by an increase in optical conductivity, were observed in LSNO (0.125). This is not solely due to the hole doping effect. Through a comprehensive analysis the experimental results, we have identified that the strength of the Ni-O orbital hybridization plays an important role in determining the effective interaction, $U^*$, within the system. This, in turn, influences the formation and dissipation of correlated plasmons in the LSNO samples. This study provides valuable insights into how orbital hybridization serves as a means to regulate correlated charge excitation and physical properties of the strongly correlated system, underscoring the compelling potential of LSNO as an alternative material within the domain of optoelectronic devices. Furthermore, it prompts further exploration into whether orbital hybridization plays a role in nickel oxide superconductivity, opening new avenues for further investigations.

## Method

**Sample Preparation**

High-quality epitaxial films of $La_{1-x}Sr_xNiO_3$, where x = 0, 0.125, and 0.25, and possessing a film thickness of 20 u.c, were synthesized on (001) LAO substrates via OPA-MBE. The deposition involved the evaporation of La, Sr, and Ni from Knudsen effusion cells, with calibrated evaporation rates monitored by a quartz crystal oscillator. The substrate temperature was set at 650 °C, and the activated oxygen partial pressure maintained at ~$5\times10^{-6}$ Torr throughout the synthesis. In-situ reflection high-energy electron diffraction (RHEED) was employed for real-time monitoring of growth rate, surface crystallography, and structure. After the growth process, the activated oxygen partial pressure was increased to $3\times10^{-5}$ Torr for an additional 30 minutes to further mitigate any residual oxygen vacancies.

**Transport Characterization and Reciprocal Space Mapping Measurements**

Electrical transport measurement as then conducted using the Keithley 2400 Source meters, and 2002 Multimeters.

To obtain the crystal structure of the LSNO films, high-resolution X-ray diffraction characterizations have been performed at the X-ray Demonstration and Development (XDD) beamline at the Singapore Synchrotron Light Source (SSLS). Reciprocal space mappings (RSMs) of the respective LSNO thin films are measured by coplanar diffraction geometry.

**Spectroscopic Ellipsometry Measurements**.

We used spectroscopic ellipsometry (SE) with a photon energy of 0.60– 4.30 eV to measure the ellipsometry parameters Ψ (the ratio between the amplitude of p- and s-polarized reflected light) and Δ (the phase difference between of p- and s-polarized reflected light) with a 70° incident angle, which indicates that the spectra include both in-plane and out-of-plane contributions. The optical conductivity has been extracted from the parameters Ψ and Δ by utilizing an air/LSNO/LAO multilayer model (see details in Supporting Information), where the LSNO composed of an average homogeneous and uniform medium. The experiments are done at 77K, 170K and room temperature. Spectroscopic Ellipsometry has no charging issue due to its photon-in−photon-out methodology. Spectroscopic ellipsometry (SE) measurements are conducted using a custom-made Variable Angle Spectroscopic Ellipsometer (VASE) of J. A. Woollam Co. Inc..

**X-ray Absorption Spectroscopy Measurements**. X-ray Absorption Spectroscopy (XAS) measurements were carried out on LSNO/LAO films at the Soft X-ray–Ultraviolet (SUV) beamline at the Singapore Synchrotron Light Source in a vacuum chamber with a base pressure of $\sim1\times10^{-9}$ mbar via the Total Electron Yield (TEY) mode. The incident X-ray was directed at the sample surfaces at a normal incident angle and grazing-incidence. All experiments were conducted at room temperature.

**The first-principles calculations**. We did the first principles calculations with the density functional theory method implemented in Vienna atomic simulation package (VASP).[60-63] The plane wave basis set with energy cutoff of 520 eV is used. The pseudopotential within the projected augmented wave (PAW) method is used.[64] We use the Monk Horst-Pack mesh of $n \times 6 \times 6$ where $n$ is about $30/l$, where is the $l$ is the length of the cell parameter in Angstrom. We use the PBE sol density functional,[65] which is optimized for computing the lattice structure in solid. In $La_{1-x}Sr_xNiO_3$ structure, no magnetic order is observed, so we use non-spin-polarized calculation. The $R\bar{3}c$ phase of the $LaNiO_3$ is assumed to be across all the Sr doping levels. which is modelled with a $2 \times 2 \times 2$ pseudo cubic 40-atom supercell with 0, 1, and 2 Sr atoms corresponding to x=0, 0.125, and 0.25. 1(2) Sr is replaced by La representing the doping level of x= 0.125(0.25). The positions of the dopants are selected to maximize the distance between them. Clustering effect is not considered in this work.

## Supporting Information

Additional raw data, experimental principles, methods details, and including analysis and fitting of SE and XAS spectra. (DOC)

## Data availability

The data that support the findings of this study are available within the article and its Supplementary Information. Additional relevant data are available from the corresponding authors upon reasonable request.

## Funding Sources

This work was supported by National Natural Science Foundation of China (Grant Nos. 52172271, 12374378, 52307026), the National Key R&D Program of China (Grant No. 2022YFE03150200), Shanghai Science and Technology Innovation Program (Grant No. 22511100200, 23511101600), the Strategic Priority Research Program of the Chinese Academy of Sciences (Grant No. XDB25000000). X. H acknowledges financial support from F.R.S. through the PDR Grants PROMOSPAN (T.0107.20). Work at PNNL was supported by the U.S. Department of Energy (DOE),

Office of Science, Basic Energy Sciences, Division of Materials Sciences and Engineering, Synthesis and Processing Science Program, under Award #10122. C.S.T. acknowledges the support from the NUS Emerging Scientist Fellowship.

## Acknowledgements

The authors would like to acknowledge the Singapore Synchrotron Light Source for providing the facility necessary for conducting the research. The Laboratory is a National Research Infrastructure under the National Research Foundation, Singapore. Any opinions, findings, and conclusions or recommendations expressed in this material are those of the author(s) and do not reflect the views of National Research Foundation, Singapore.

## Author contributions

X.Y. conceived the project. M.S. and X.H. contributed equally to this work. J.L., Y.D., and L.W. synthesized the samples. L.W. performed the electron transport experiments. C.S.T. performed the XRD and XAS experiments and analyzed the data with M.S. and X.Y.. M.S., C.S.T., X.L., M.C., L.D. and X.Y. performed the SE experiments and analyzed the data. X.H. performed the DFT analysis. M.S., C.S.T., X.H., L.W., and X.Y. wrote the manuscript, with input from all the authors.

## Competing interests

The authors declare no competing interests.

**Correspondence** and requests for materials should be addressed to Chi Sin Tang or Le Wang or Xinmao Yin.

020406.
(50) Stavitski, E.; de Groot, F. M. The Ctm4xas Program for Eels and XAS Spectral Shape Analysis of Transition Metal L Edges. *Micron* **2010,** *41* (7), 687-694.
(51) Guo, E. J.; Liu, Y.; Sohn, C.; Desautels, R. D.; Herklotz, A.; Liao, Z.; Nichols, J.; Freeland, J. W.; Fitzsimmons, M. R.; Lee, H. N. Oxygen Diode Formed in Nickelate Heterostructures by Chemical Potential Mismatch. *Adv. Mater.* **2018,** *30* (15), 1705904.
(52) Wang, L.; Chang, L.; Yin, X.; You, L.; Zhao, J.-L.; Guo, H.; Jin, K.; Ibrahim, K.; Wang, J.; Rusydi, A.; Wang, J. Self-Powered Sensitive and Stable Uv-Visible Photodetector Based on $GdNiO_3$/Nb-Doped $SrTiO_3$ Heterojunctions. *Appl. Phys. Lett.* **2017,** *110* (4), 043504.
(53) Ding, H.; Park, K.; Gao, Y. Evolution of the Unoccupied States in Alkali Metal-Doped Organic Semiconductor. *J. Electron Spectrosc. Relat. Phenom.* **2009,** *174* (1-3), 45-49.
(54) Winkler, S.; Amsalem, P.; Frisch, J.; Oehzelt, M.; Heimel, G.; Koch, N. Probing the Energy Levels in Hole-Doped Molecular Semiconductors. *Mater. Horiz.* **2015,** *2* (4), 427- 433.
(55) West, P. R.; Ishii, S.; Naik, G. V.; Emani, N. K.; Shalaev, V. M.; Boltasseva, A. Searching for Better Plasmonic Materials. *Laser Photonics Rev.* **2010,** *4* (6), 795-808.
(56) Olmon, R. L.; Slovick, B.; Johnson, T. W.; Shelton, D.; Oh, S.-H.; Boreman, G. D.; Raschke, M. B. Optical Dielectric Function of Gold. *Phys. Rev. B* **2012,** *86* (23), 235147.
(57) Jablan, M.; Soljacic, M.; Buljan, H. Plasmons in Graphene: Fundamental Properties and Potential Applications. *Proc. IEEE* **2013,** *101* (7), 1689-1704.
(58) Goldflam, M. D.; Liu, M. K.; Chapler, B. C.; Stinson, H. T.; Sternbach, A. J.; McLeod, A. S.; Zhang, J. D.; Geng, K.; Royal, M.; Kim, B.-J.; Averitt, R. D.; Jokerst, N. M.; Smith, D. R.; Kim, H. T.; Basov, D. N. Voltage Switching of a $VO_2$ Memory Metasurface Using Ionic Gel. *Appl. Phys. Lett.* **2014,** *105* (4), 041117.
(59) Markov, P.; Appavoo, K.; Haglund, R. F., Jr.; Weiss, S. M. Hybrid Si-VO2-Au Optical Modulator Based on near-Field Plasmonic Coupling. *Opt. Express* **2015,** *23* (5), 6878-87.
(60) Kresse, G.; Furthmiiller, J. Efficiency of Ab-Initio Total Energy Calculations for Metals and Semiconductors Using a Plane-Wave Basis Set *Comput. Mater. Sci.* **1996,** *6* (1), 15-50.
(61) Kresse, G.; Furthmüller, J. Efficient Iterative Schemes for Ab Initio Total-Energy Calculations Using a Plane-Wave Basis Set. *Phys. Rev. B* **1996,** *54* (16), 11169-11186.
(62) Kresse, G.; Hafner, J. Ab Initio Molecular Dynamics for Liquid Metals. *Phys. Rev. B* **1993,** *47* (1), 558-561.
(63) Kresse, G.; Hafner, J. Ab Initio Molecular-Dynamics Simulation of the Liquid-Metal —Amorphous-Semiconductor Transition in Germanium. *Phys. Rev. B* **1994,** *49* (20), 14251-14269.
(64) Kresse, G.; Joubert, D. From Ultrasoft Pseudopotentials to the Projector Augmented-Wave Method. *Phys. Rev. B* **1999,** *59* (3), 1758-1775.
(65) Perdew, J. P.; Ruzsinszky, A.; Csonka, G. I.; Vydrov, O. A.; Scuseria, G. E.; Constantin, L. A.; Zhou, X.; Burke, K. Restoring the Density-Gradient Expansion for Exchange in Solids and Surfaces. *Phys. Rev. Lett.* **2008,** *100* (13), 136406.